\documentclass[12pt]{iopart}

\usepackage{graphicx}
\usepackage{epstopdf}

\begin{document}

\title[]{Study of internal structures of ${}^{9,10}$Be and ${}^{10}$B in scattering of ${}^{4}$He from ${}^{9}$Be}

\author{S~M~Lukyanov$^1$, A~S~Denikin$^{2,1}$, E~I Voskoboynik$^1$, S~V~Khlebnikov$^{3,4}$, M~N~Harakeh$^{5,6}$, V~A~Maslov$^1$, Yu~E~Penionzhkevich$^1$, Yu~G~Sobolev$^1$, W~H~Trzaska$^3$, and G~P~Tyurin$^3$ and K~A~Kuterbekov$^{7}$}
\address{$^1$ Flerov Laboratory of Nuclear Reactions, Dubna, Russian Federation}
\address{$^2$ International University "Dubna", Dubna, Russian Federation}
\address{$^3$ Department of Physics, University of Jyv\"askyl\"a, Jyv\"askyl\"a, Finland}
\address{$^4$ Khlopin Institute, St. Petersburg, Russian Federation}
\address{$^5$ Kernfysisch Versneller Instituut, University of Groningen, Groningen, the Netherlands}
\address{$^6$ GANIL, CEA/DSM-CNRS/IN2P3, 14076 Caen, France}
\address{$^7$  Eurasian Gumilev University, Astana, Kazakhstan}

\ead{lukyan@nrmail.jinr.ru}
\begin{abstract}
A study of inelastic scattering and single-particle transfer reactions was performed by an alpha beam at 63 MeV on a ${}^{9}$Be target. Angular distributions of the differential cross sections for the ${}^{9}$Be($\alpha$,$\alpha'$)${}^{9}$Be$^*$, ${}^{9}$Be($\alpha, {}^{3}$He)${}^{10}$Be and ${}^{9}$Be($\alpha$,t)${}^{10}$B reactions were measured. Experimental angular distributions of the differential cross sections for the ground state and a few low-lying states were analyzed in the framework of the optical model, coupled channels and distorted-wave Born approximation. An analysis of the obtained spectroscopic factors was performed.
\end{abstract}

\submitto{\JPG}
\maketitle

\section{Introduction}
In recent years, the study of light, weakly-bound nuclei \cite{Ref1,Ref2} has intensified due to the significant progress made with radioactive beam facilities. It has led to a resurgence of interest in the study of light stable nuclei such as ${}^{6,7}$Li and ${}^{9}$Be, for example. It has been shown that in light nuclei the nucleons tend to group into clusters, whose relative motion mainly defines the properties of these nuclei. Consequently, the cluster structures of their ground as well as low-lying excited states have been in the focus of studies. As examples, nuclei ${}^{6}$Li and ${}^{7}$Li are both well described by two-body cluster models ($\alpha$+d and $\alpha$+t, respectively). Another interesting nuclide is ${}^{9}$Be, which could be described as an $\alpha$+$\alpha$+n three-body configuration; one may also consider it as a nuclear system with two-body configuration ${}^{8}$Be+n or ${}^{5}$He+$\alpha$.

The addition of a second valence neutron to ${}^{9}$Be leads to another intriguing nucleus, ${}^{10}$Be. A microscopic $\alpha$+$\alpha$+n+n cluster model was proposed for ${}^{10}$Be in order to clarify the relation between the configurations of the valence neutrons and the $\alpha$+$\alpha$ core. In spite of its large binding, the $\alpha$-$\alpha$ clustering in the ground state persists due to a coupling effect between the ${}^{6}$He+$\alpha$ and the ${}^{5}$He+${}^{5}$He configurations.

Recently, special attention has been focused on the role of the extra "valence" nucleons, and their influence on the cluster structure of the excited states \cite{Ref3}. A semi-quantitative discussion of this subject can be found in Ref.\cite{Ref3}, where the two-center molecular states in ${}^{9}$B, ${}^{9}$Be, ${}^{10}$Be, and ${}^{10}$B nuclei were considered in the framework of a molecular-type model.

One of the tools to study nuclear structure is scattering of a projectile, such as p or ${}^{3,4}$He, from a target nucleus, the structure of which is going to be studied. This method is based on angular-distribution measurements of elastic and inelastic scattering of projectile-like products. The energy spectra of these products bear information about the internal structure of the incoming and outgoing nuclei.

Alpha scattering from ${}^{9}$Be target at E$_\alpha$ = 65 MeV was measured in details for the first time in Ref.\cite{Ref4} and later in Ref.\cite{Ref5}. Optical model analysis of the elastic scattering data was performed, and distorted-wave Born approximation (DWBA) and coupled-channel (CC) calculations were also done for inelastic scattering and single-particle transfer channels. A molecular-type rotational band was used to describe the data.

The data of Ref.\cite{Ref6} demonstrated the measurement in inverse kinematics of 65 MeV ${}^{12}$C beam scattered from a ${}^{9}$Be target. The experimental data were analyzed in Ref.\cite{Ref7}. The calculations agree well with data for elastic scattering and excitation of the 5/2$^-$ resonance of ${}^{9}$Be at 2.43 MeV while data on the 1/2$^+$ (1.68 MeV) state excitation was not well described by their model. This was as expected from structure calculations of ${}^{9}$Be treating this state as almost pure ${}^{8}$Be+n cluster configuration. A rather different conclusion was drawn in Ref.\cite{Ref8}. It was found that the decay branch n+${}^{8}$Be(2$^+$) provides a small fraction of the decay of 5/2$^-$ state. In total agreement with this finding, Charity et al. \cite{Ref9} recently confirmed that decay of ${}^9$B has a dominant branch to $\alpha$+${}^5$Li implying that "the corresponding mirror state in ${}^{9}$Be would be expected to decay through the mirror channel $\alpha$ + ${}^{5}$He, instead of through the n + ${}^{8}$Be(2$^+$) channel."

This article is an attempt to shed light on the internal structure of ${}^{9,10}$Be and ${}^{10}$B nuclei by (in)elastic scattering of ${}^{4}$He ions on ${}^{9}$Be target. We expected that the sensitivity of the high precision alpha scattering data to the cluster structure of ${}^{9}$Be could be demonstrated.

\section{Experiment}
The experiment was performed at the K130 Cyclotron facility of the Accelerator Laboratory of the Physics Department of Jyv\"{a}skyl\"{a} University. The beam energy of ${}^{4}$He ions was 63 MeV. The average beam current during the experiment was maintained at 3 nA. The self-supporting Be target was prepared from a 99\% pure thin foil of beryllium. The target thickness was 7 $\mu$m. Peaks due to carbon and oxygen contaminations were not observed in the energy spectra.

To measure (in)-elastically scattered ions, two telescopes each consisting of Si-Si(Li) detectors with thicknesses of 100 $\mu$m and 3 mm, respectively, were used. Each pair of detectors was mounted at a distance of about 45 cm from the target. Particle identification was performed based on the energy-loss measurements of $\Delta$E and residual energy E$_r$, i.e. the so-called $\Delta$E-E method. The Si-telescopes were mounted on rotating supports, which allowed to obtain data from $\theta_{lab}$ = 20$^\circ$ to $\theta_{lab}$ = 107$^\circ$ in steps of 1-2$^\circ$.

\begin{figure}[t]
\begin{center}\includegraphics[width=250pt]{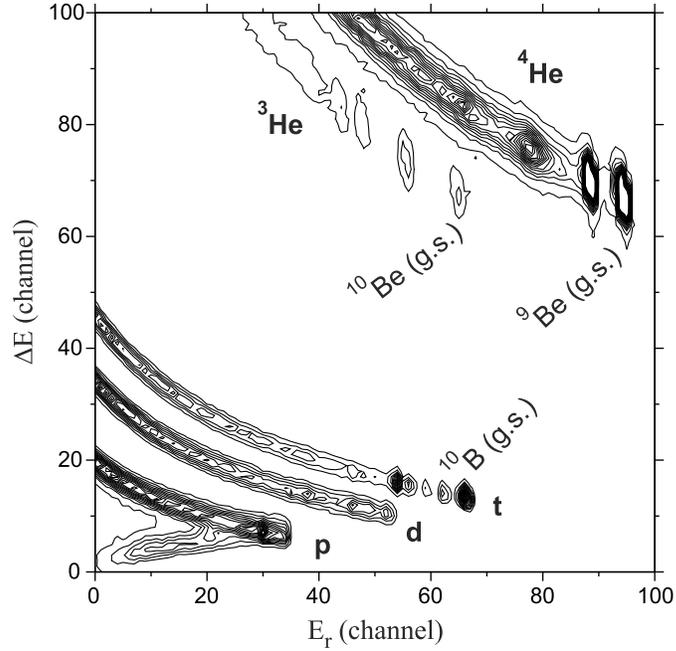}\end{center}
\caption{Reaction product yields versus measured energy loss $\Delta$E and residual energy E$_r$ measured by the Si-Si(Li) telescope. The loci for ${}^{3,4}$He, p, d and t are visible.}\label{fig1}
\end{figure}

The overall energy resolution of the telescopes was nearly 200 keV. An example of two-dimensional plot (yield versus energy loss $\Delta$E and residual energy E$_r$, measured by Si-Si(Li) detectors) is shown in Fig.\ref{fig1}. Excellent energy resolutions of both $\Delta$E and E detectors allowed identifying ${}^{3,4}$He, t, d and p unambiguously. These detected particles were produced in the channels listed in Table \ref{Table1}.
\Table{\label{Table1}Reaction channels and their Q- values}
\br
Reaction channel & Q-values (MeV)\\
\mr
${}^{9}$Be + ${}^{4}$He $\to$ ${}^{10}$Be + ${}^{3}$He & $-$13.8 \\
${}^{9}$Be + ${}^{4}$He $\to$ ${}^{10}$B  + t   & $-$13.2 \\
${}^{9}$Be + ${}^{4}$He $\to$ ${}^{11}$B  + d   & $-$8.0 \\
${}^{9}$Be + ${}^{4}$He $\to$ ${}^{12}$B  + p   & $-$6.9 \\
\br
\endTable

The channel leading to the production of ${}^{7}$Be+${}^{6}$He has minimal probability due to the low Q-value. Other reaction channels take places at higher Q-values and consequently have larger cross sections, as it is shown in Fig.\ref{fig1}. The production yield of ${}^{6}$He starts to be visible only when plotting the z-axis (yield) in logarithmic scale; it is not shown in Fig.\ref{fig1}.

\begin{figure}[t]
\begin{center}\includegraphics[width=300pt]{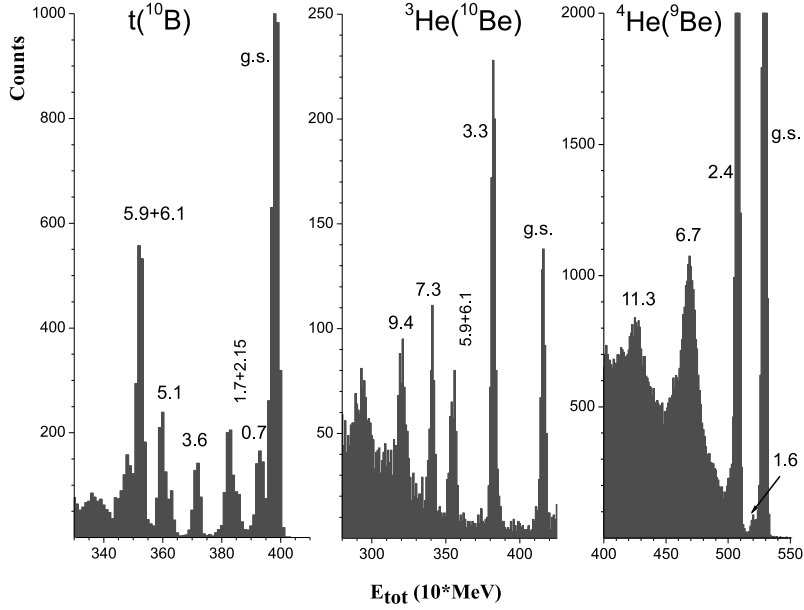}\end{center}
\caption{Measured spectra of total energies for ${}^{9}$Be($\alpha$,t)${}^{10}$B (left), ${}^{9}$Be($\alpha$,${}^{3}$He)${}^{10}$Be (middle) and ${}^{9}$Be($\alpha$, $\alpha'$)${}^{9}$Be (right) reaction channels. The ground and most populated excited states of ${}^{10}$B, ${}^{10}$Be and ${}^{9}$Be are unambiguously identified.}\label{fig2}
\end{figure}

Comparing with the experimental technique of Ref.\cite{Ref5} we have the advantage to distinguish the particles p, d, t, ${}^{3}$He and ${}^{4}$He and determine their total deposited energies. The total energies were obtained after energy calibration of all Si-detectors and summing of energy deposits in the $\Delta$E and E$_r$ detectors. The spectra of total deposited energy are shown in Fig.\ref{fig2} for t, ${}^{3}$He and ${}^{4}$He. All peaks, which can be observed in the histograms in Fig.\ref{fig2}, were identified and found to belong to the ground and excited states of ${}^{9}$Be, ${}^{10}$Be and ${}^{10}$B, as the complementary products to detected particles ${}^{4}$He, ${}^{3}$He, and t, respectively. We were not able to get information about states in ${}^{11,12}$B, due to the restricted thickness of the Si(Li) detectors, as a consequence of which p and d punched through and were not stopped in the Si(Li) detectors.

We found excellent agreement between excited states observed in our experiment with those previously measured for ${}^{9}$Be \cite{Ref4, Ref5, Ref10}, ${}^{10}$Be \cite{Ref5, Ref11}, and ${}^{10}$B \cite{Ref5, Ref12}. Because the incident beam energy was rather high (15.75 MeV/u), the observed states are most likely populated in one-step direct transfer reactions. Another advantage is that the two nuclei ${}^{10}$Be and ${}^{10}$B, belonging to A = 10 multiplet, are populated in the same reaction.

\section{Results}

\subsection{${}^{9}$Be} 
\begin{figure}[t]
\begin{center}\includegraphics[width=200pt]{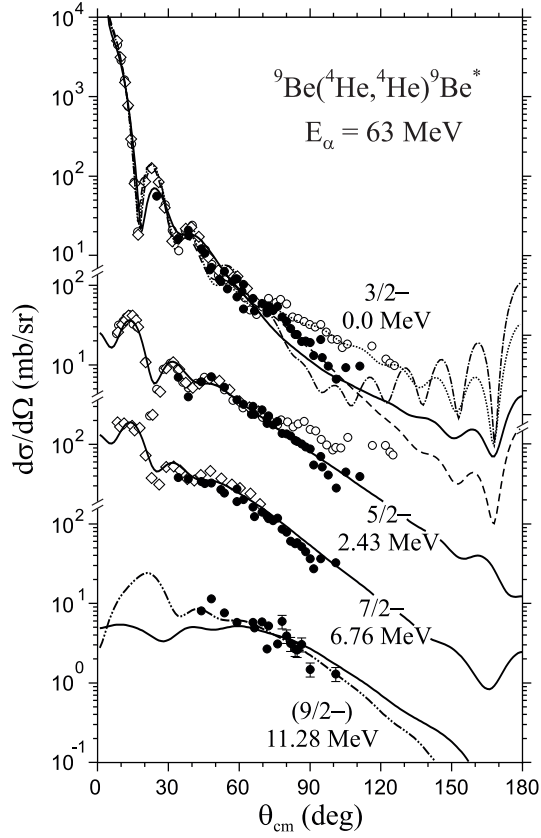}\end{center}
\caption{Differential cross sections for the ${}^4$He (63 MeV) + ${}^9$Be scattering reactions. Data obtained in the present work are shown by symbols $\fullcircle$, data from Refs. [4,5] are denoted by $\opendiamond$ and $\opencircle$ open symbols, respectively. Angular distributions of elastic and inelastic scattering to the 5/2$^-$ (2.43 MeV), 7/2$^-$ (6.76 MeV) and 9/2$^-$ (11.28 MeV) ${}^9$Be states are shown. The curves represent the results of the optical-model and coupled-channel calculations (see explanation in text).}\label{fig3}
\end{figure}

Measured differential cross sections of the ground and low-lying excited states for ${}^{9}$Be are presented in Fig.\ref{fig3}. Due to low statistics we were not able to get the angular distribution for the first-excited 1/2$^+$ state of ${}^{9}$Be at 1.6 MeV. The oscillations at small angles of the ground state (3/2$^-$) and first-excited state (5/2$^-$) are in anti-phase. No significant oscillatory structure was observed for the angular distributions of the 7/2$^-$ and 9/2$^-$ states.

Comparison with the results for the ground state of the previous measurements \cite{Ref4,Ref5} (open symbols) demonstrates a good agreement at small scattering angles. The disagreement is observed at angles larger than 70$^\circ$ where our data are smaller than those of Ref.\cite{Ref5}. From the technical point of view, this difference could be explained by absence of particle identification in Ref.\cite{Ref5} where Si(Li) detectors were used to measure the total energy only, without a $\Delta$E measurement that would allow Z and A identification of the detected particles. Another reason could also be due to a different method used for subtraction of the continuum under the peak. The same reasons are responsible for the difference between our data and those of Ref.\cite{Ref5} for the level at 6.76 MeV in the angular range 30-60 degrees (see Fig.\ref{fig3}).

Fig.\ref{fig3} shows measured differential cross sections for the elastic and inelastic scattering (symbols) together with the results of theoretical calculations (curves) performed within optical model (OM) and coupled-channel (CC) approach. Theoretical curves were obtained with the aid of NRV server optical model routine \cite{Ref13} and the ECIS06 coupled-channel code \cite{Ref14,Ref15}.

Firstly, let us consider the analysis of the elastic scattering cross section. The optical potential was chosen in the usual Woods-Saxon form
$$V(r) =  - {V_0}f(r,{R_V},{a_V}) - i{W_0}f(r,{R_W},{a_W}),$$
where the function $f(r,R,a) = {(1 + {e^{(r - R)/a}})^{ - 1}}$. Potential parameters (OM1) fitted within the optical model to the measured experimental data are listed in Table \ref{Table2}. Corresponding curve is shown in Fig.\ref{fig3} as a dashed line and demonstrates good agreement with obtained data. In addition, the calculation of elastic scattering cross section was performed with the parameters (OM2), recommended in Ref.\cite{Ref4}. It is plotted as dash-dotted line in Fig.\ref{fig3}. One may see that OM2 parameters of Ref.\cite{Ref4} provide poorer agreement with data at large angles since the OM2 parameters were obtained by fitting the data in narrower angular range. The main differences between the OM1 and OM2 sets are the shallower depth of real part and the larger diffuseness parameters of the potential found in this work.

Our fitting (OM3) of the data \cite{Ref5} for the ground state of ${}^{9}$Be are also given in Table \ref{Table2}. Corresponding angular distribution is shown in Fig.\ref{fig3} by dotted line. Obtained parameters are rather close to the ones found in Ref. \cite{Ref5}, except noticeably smaller radius of the imaginary part. It results in the larger elastic scattering cross sections at large deflection angles.
\begin{table}
\caption{\label{Table2}Potential parameters used within optical model and coupled channels approaches}
\footnotesize\rm
\begin{tabular*}{\textwidth}{@{}l*{15}{@{\extracolsep{0pt plus12pt}}c}}
\br
             & $V_0$(MeV) & $r_V$(fm) & $a_V$ (fm) & $W_0$(MeV) & $r_W$(fm) & $a_W$ (fm) & $r_C$(fm)\\
\mr
$\alpha$ + ${}^{9}$Be (OM1)           & 101.0 & 1.40 & 0.75 & 32.70 & 1.50 & 0.75 & 1.30 \\
$\alpha$ + ${}^{9}$Be (CC)            & 96.82 & 1.19 & 0.75 & 11.84 & 1.61 & 0.75 & 1.30 \\
$\alpha$ + ${}^{9}$Be (OM2)\cite{Ref4}& 146.8 & 1.49 & 0.59 & 35.95 & 1.49 & 0.59 & 1.25 \\
$\alpha$ + ${}^{9}$Be (OM3)$^{\rm a}$ & 109.1 & 1.48 & 0.65 & 40.00 & 1.34 & 0.727 & 1.30 \\
${}^3$He + ${}^{10}$Be$^{\rm b}$      & 95.0 & 0.95 & 0.815 & 8.00 & 1.60 & 0.726 & 1.07 \\
${}^3$H  + ${}^{10}$B$^{\rm c}$       & 95.0 & 1.04 & 0.815 & 3.00 & 1.867 & 0.47 & 1.07 \\
${}^{3}$He or ${}^3$H + ${}^{10}$B \cite{Ref4}& 132.9 & 1.54 & 0.57 & 19.50 & 1.82 & 0.22 & 0.81 \\
\br
\end{tabular*}
\noindent $^{\rm a}$ {Parameters are obtained by fitting the elastic scattering data of Ref.\cite{Ref5}.}\\
\noindent $^{\rm b}$ {Parameters were taken from Ref.\cite{Ref28}. Real and imaginary depths and radii were modified within 10-15\% of magnitude in order to fit experimental data on transfer to the ${}^{10}$Be ground state.}\\
\noindent $^{\rm c}$ {The ${}^{3}$He+${}^{10}$Be parameters (were taken from Ref.\cite{Ref28}) were used as initial set and then parameters were fitted to reproduce experimental data on transfer to the ${}^{10}$B ground state.}
\end{table}

The simplest view of the ${}^{9}$Be nucleus is that it is a strongly deformed three-body system  consisting of two $\alpha$ particles held together by a weakly bound neutron. It is very natural that different molecule-like states may appear in the excited states. It is the aim of research in this mass region to make a systematic study of structure changes with increasing excitation energy.

Due to the Borromean structure of ${}^{9}$Be, it will be configured as two alpha particles plus a neutron or as two unstable intermediate nuclei: (i) ${}^{8}$Be or (ii) ${}^{5}$He in combination with a neutron and an $\alpha$-particle, respectively. However, to distinguish break-up into ${}^{4}$He and ${}^{5}$He is not a trivial kinematical problem; nevertheless, some attempt has been successfully undertaken \cite{Ref16}. The structure of ${}^{9}$Be through ${}^{8}$Be+n has been quantified for the low-lying excited states in ${}^{9}$Be. Higher excited levels are associated with a ${}^{5}$He cluster. An aim of the present experiment was to study the peculiarity of the angular distributions of elastic and inelastic scattering, mainly for 5/2$^-$, 7/2$^-$ and 9/2$^-$ states, to try to learn something about their cluster structure.

Analysis of inelastic scattering data within the DWBA or CC approach allows to extract the information on the deformation of an excited nucleus treating these states as collective rotational excitations. Corresponding coupling matrix elements in addition to the radial form-factor includes the deformation length $\beta_\lambda R_V$, where quantity $\beta_\lambda$ is a deformation parameter, $\lambda$ is a multipolarity of the transition defined by the transferred angular momentum and $R_V = r_V A^{1/3}$ is an interaction radius depending on the mass A of excited nucleus.

It is known \cite{Ref4, Ref5, Ref17, Ref18, Ref19} that ${}^{9}$Be has a rotational band ($K^\pi$ = 3/2$^-$) built on its ground state. In previous studies only ground and excited states of the band were analyzed together in the CC framework. One may expect \cite{Ref21_1, Ref21_2} that all angular distributions shown in Fig.\ref{fig3} are related to the same rotational band. So far, the values of spin and parity of the 11.28 MeV state were uncertain. No direct measurements were done. This level was listed either 7/2$^-$ or 9/2$^-$ state in the literature and databases. Following Ref.\cite{Ref18}, we consider this state to belong to the rotational band and therefore to have spin-parity 9/2$^-$ (see data at bottom of Fig.\ref{fig3} and further explanation of the curves below).

The solid lines in Fig.\ref{fig3} represent the results of a CC calculation within the symmetric rotational model taking also into account  Coulomb excitation and reorientation terms. The ECIS06 code was employed. The parameters of the optical potential used in the CC calculations are given in Table \ref{Table2}. They were fitted to the data shown in Fig.\ref{fig3}, using the OM1 parameters as an initial set. It was found that inelastic scattering data for the first three states of the rotational band may be well described if one assumes  $\beta_\lambda R_V$ = 1.574 fm and $\beta_2$ = 0.64. These values are consistent with results of previous studies \cite{Ref4}.

Quadrupole moment Q$_{20}$ of the ${}^{9}$Be nucleus is known to be equal +53 mb \cite{Ref20_1,Ref20_2} indicating a prolate deformation for the ground state. Previous studies (e.g. Refs. \cite{Ref4, Ref5}) have shown a quite large deformation parameter $\beta_2$ lying in the range 0.5 to 0.7. It provided rather good agreement with our data on elastic and inelastic (2.43 MeV and 6.76 MeV states) scattering. The obtained large $\beta_2$ value may be considered as the confirmation of the cluster structure of the low-lying states of ${}^{9}$Be. However it doesn't allow to give unambiguous preference to the one of the possible configurations, for example, ($\alpha$+$\alpha$+n) or ($\alpha$+${}^{5}$He).

In Fig.\ref{fig3}, one may see rather good agreement between CC calculation and the experimental data (see solid line in the bottom part of Fig.\ref{fig3}) in the case of 11.28 MeV state. In order to improve the fits for this state, an additional hexadecapole term $\beta_4$ in the definition of the ${}^{9}$Be radius was added. The dash-double-dotted line in Fig.\ref{fig3} demonstrates the result obtained with the same $\beta_2$ value and $\beta_4$ = 0.27, which agrees much better with the data. There is insignificant influence of the $\beta_4$ parameter on the cross sections for the 3/2$^-$,  5/2$^-$ and 7/2$^-$ states. This may be evidence of the different structure of the 9/2$^-$ state of the ${}^{9}$Be nucleus. It should be noted that data on inelastic scattering to the 11.28 MeV state were measured in the middle range of the angles, where the  two  theoretical predictions are rather comparable. Thus, in order to draw final conclusion, additional measurements are required in a broader region of the scattering angles.

\begin{figure}[h]
\begin{center}\includegraphics[width=200pt]{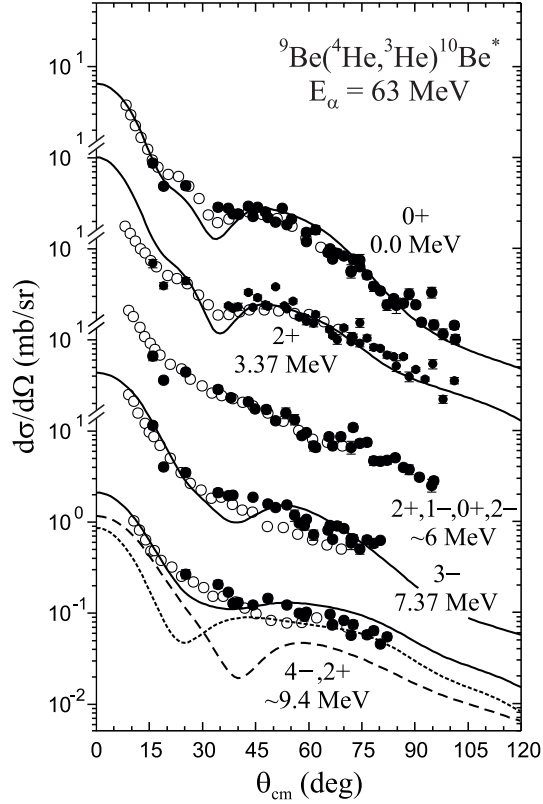}\end{center}
\caption{Angular distributions of the differential cross sections for the ground and low-lying excited states for ${}^{10}$Be in the reaction ${}^9$Be(${}^4$He,${}^3$He)${}^{10}$Be: the ground state 0$^+$ of ${}^{10}$Be, the first 2$^+$ state at 3.4 MeV, sum of cross sections for 2$^+$, 1$^-$, 0$^+$ and 2$^-$ levels between 5.9 and 6.3 MeV, the 3$^-$ level at 7.3 MeV and for 4$^-$ and 2$^+$ levels at 9.3 MeV. Results of the present experiment are shown by solid symbols. Data from Ref.\cite{Ref4} are presented by open symbols. No DWBA curve is drawn through the differential cross sections for the 2$^+$, 1$^-$, 0$^+$ and 2$^-$ states, which are very close in the energy and therefore could not be resolved. The curves are explained in the text.}\label{fig4}
\end{figure}

\subsection{${}^{10}$Be}
If ${}^{9}$Be shows molecular cluster structure \cite{Ref3}, then ${}^{10}$Be might be expected to show more sophisticated internal structure. Molecular structure of the ${}^{10}$Be nucleus is formed by two alpha particles and two neutrons. Such constitution attracts even more interest, since one neutron added to ${}^{9}$Be makes the ${}^{10}$Be nucleus tightly bound \cite{Ref15}.

In this work we performed measurements of the angular distributions for the ${}^{9}$Be(${}^{4}$He,${}^{3}$He)${}^{10}$Be reaction, leading to different ${}^{10}$Be excited states. Angular distributions of the differential cross sections for the ground and low-lying excited states for ${}^{10}$Be are plotted in Fig.\ref{fig4}. Results of the present experiment are shown by solid symbols; data from Ref.\cite{Ref4} are presented by the open symbols. Solid lines are the result of the finite-range DWBA calculations with the DWUCK5 code \cite{Ref24}. This type of the calculation is available via the internet web page of the NRV project \cite{Ref23}.

In order to perform the DWBA calculations the OM1 parameters for the entrance channel and the corresponding potential for the exit channel were chosen to calculate distorted waves (see Table \ref{Table1}). OM parameters for the exit channel ${}^{3}$He + ${}^{10}$Be were chosen close to the potential recommended in Ref.\cite{Ref28}. According to Table \ref{Table1}, Q for the ${}^{9}$Be(${}^{4}$He, ${}^{3}$He)${}^{10}$Be reaction channel is negative with a large absolute value. It legitimizes a slight variation of  the optical model parameters for the exit channel (within 10\%) for better agreement of the calculations with the data. In the analysis reported below we varied only the depths of the real and imaginary parts within indicated limits.

The single-particle wave functions in the entrance and exit channels were defined within standard potential model \cite{Ref26,Ref27}. The interaction for n + ${}^{3}$He system was chosen of the Gaussian form
$$V(r) =  - {V_G}\exp \left( -\frac{r^2}{R_G^2}\right),$$
where the radius $R_G$ = 2.452 fm \cite{Ref27}, while the potential depth $V_G$ is fitted to reproduce the correct value of neutron binding energy E$_n$ = $-$20.58 MeV in the ${}^{4}$He nucleus. The n + ${}^{9}$Be potential in the final state was defined as a real Woods-Saxon potential with radius $R_V$ = 1.26 A$_\mathrm{Be}^{1/3}$ fm and diffuseness $a_V$ = 0.6 fm. Potential depth $V_0$ defined in the same manner as $V_G$ parameter. For states unbound to the neutron emission in ${}^{10}$Be, the single particle was assumed to be bound by 0.1 MeV, as it was suggested in \cite{Ref4}.

Relative angular momentum of neutron state in the projectile or target-like fragment was fixed by the total momentum $J$ and parity $\pi$ conservation laws. In particular, the ground state of the ${}^{10}$Be(0$^+$) = n(1/2$^+$) + ${}^{9}$Be(3/2$^-$) nucleus was considered as 1p$_{3/2}$ neutron state, while the excited states of ${}^{10}$Be with negative parity was treated as 1d$_{5/2}$ neutron state. All spectroscopic properties of the ${}^{10}$Be excited states are listed in Table \ref{Table3}.

The DWBA differential cross section for the considered stripping reactions can be compared with experimental data in the following way \cite{Ref24}
$$\frac{{d{\sigma _{\exp }}}}{{d\Omega }} = {S_i}{S_f}\frac{{(2{J_f} + 1)}}{{(2{J_i} + 1)}}{\sigma _{DW}}(q),$$
where $J_i$ = 3/2 and $J_f$ are the angular momenta of the ${}^{9}$Be target and the final state populated in ${}^{10}$Be, respectively, $\sigma_{DW}(\theta)$ is the output from DWUCK5, S$_i$ and S$_f$ are the projectile and target-like fragment spectroscopic factors, respectively. Fig.\ref{fig4} demonstrates how good are the obtained absolute values of the spectroscopic factors S$_f$, which were obtained from the comparison of the measured angular distributions and DWUCK5 calculations for the different ${}^{10}$Be final states. The values of the spectroscopic factors are listed in Table \ref{Table3} together with S$_f$ reported in Ref.\cite{Ref4}.

\begin{table}
\begin{center}
\caption{\label{Table3}Spectroscopic information for the ${}^{9}$Be(${}^{4}$He,${}^{3}$He)${}^{10}$Be and ${}^{9}$Be(${}^{4}$He,t)${}^{10}$B reactions as obtained from the DWBA analysis.
${}^{9}$Be($\alpha$,${}^{3}$He)${}^{10}$Be	${}^{9}$Be($\alpha$,t)${}^{10}$B}
\footnotesize\rm
\begin{tabular*}{\textwidth}{@{}l*{15}{@{\extracolsep{0pt plus12pt}}c}}
\br
&&${}^{9}$Be($\alpha$,${}^{3}$He)${}^{10}$Be&&&&& ${}^{9}$Be($\alpha$,t)${}^{10}$B&&\\
\mr
E$_x$,(MeV) & J$\pi$ & $l$ & S$_f$ \cite{Ref4} & S$_f$, present & E$_x$ (MeV) & J$\pi$ & $l$ & S$_f$ \cite{Ref4} & S$_f$, present \\
\mr
g.s.  & 0$^+$ & 1 & 1.58      & 1.65      & g.s. & 3$^+$ & 1 & 0.89      & 0.59 \\
3.368 & 2$^+$ & 1 & 0.38      & 1.00      & 0.781& 1$^+$ & 1 & 1         & 1.0 \\
5.958 & 2$^+$ & 1 & $\le$0.73 & $\le$1.40 & 1.76 & 0$^+$ & 1 & 1.58      & 1.38 \\
5.960 & 1$^-$ & 2 & $\le$0.14 & $\le$0.43 & 2.1  & 1$^+$ & 1 & 0.52      & 0.30 \\
6.179 & 0$^+$ & 1 & --        & --        & 3.6  & 2$^+$ & 1 & 0.28      & 0.23 \\
6.263 & 2$^-$ & 2 & 0.08      & $\le$0.26 & 5.11 & 2$^-$ & 2 & $\le$0.27 & $\le$0.16 \\
7.371 & 3$^-$ & 2 & 0.26      & 0.28      & 5.16 & 2$^+$ & 1 & $\le$1.85 & $\le$0.75 \\
7.542 & 2$^+$ & 1 & --        & --        & 5.18 & 1$^+$ & 1 & $\le$3.14 & $\le$1.0 \\
9.27  &(4$^-$)& 2 & $\le$0.18 & 0.10      & 5.93 & 2$^+$ & 1 & 0.48      & $\le$0.95 \\
9.56  & 2$^+$ & 1 & --        & 0.23      & 6.13 & 3$^-$ & 2 & 0.24      & $\le$0.19 \\
\br
\end{tabular*}
\end{center}
\end{table}

It is seen that obtained cross sections agree well with data. The spectroscopic factors extracted from our analysis are very close to the ones listed in Ref.\cite{Ref4} (see Table \ref{Table3}), except for the ${}^{10}$Be state at 3.368 MeV, where spectroscopic values differ by more a factor two. The reason for this discrepancy is the following. The spectroscopic factor in our work was defined by adjusting the theoretical curve to the data measured in middle angle domain, while in Ref.\cite{Ref4} it was fitted to the forward experimental points near $\theta_{cm} \approx$ 10$^\circ$.

Because of not-optimal value of the energy resolution, we were not able to separate the excited states nearby 6 MeV. Low statistics did not allow to observe the 2$^+$ state at 7.54 MeV. The experimental cross sections corresponding to two overlapping states near 9.5 MeV were described as a sum of DWUCK5 outputs multiplied by the corresponding spectroscopic factors. Table \ref{Table3} contains the S$_f$ values providing the best fit. Short and long-dashed lines in Fig.\ref{fig4} show the contribution from the 2$^+$ and 4$^-$ states, respectively.

Note that in order to describe the data one needs much smaller radius of the real part of the optical potential for the exit channel ($r_V$ = 0.95 fm) in comparison to the radius in the entrance channel ($r_V$ = 1.40 fm). This could be interpreted as due to the compactness of the ${}^{10}$Be nucleus.

Nuclear charge radii of ${}^{7,9,10}$Be have been measured by high precision laser spectroscopy \cite{Ref21_1, Ref21_2}: the charge radius decreases form ${}^{7}$Be to ${}^{10}$Be. Comparing the Coulomb parameter $r_C$ with that of ${}^{9}$Be, we obtained a smaller value of $r_C$ for ${}^{10}$Be. In Ref.\cite{Ref21_1, Ref21_2}, the decrease was explained as probably caused by the clusterization of ${}^{7}$Be into an $\alpha$ and triton clusters, whereas ${}^{9,10}$Be were considered to be $\alpha$+$\alpha$+n and $\alpha$+$\alpha$+n+n systems, respectively, and were more compact. The experimental trend was shown \cite{Ref15}, to change beyond ${}^{10}$Be with an increase of the charge radius with atomic mass. Furthermore, the large experimental value of the charge radius for ${}^{12}$Be is consistent with a breakdown of the N = 8 shell closure.

The root-mean-square matter radii deduced by means of Glauber-model analysis with an optical limit approximation were reported in Ref.\cite{Ref22} and didn't show large difference in values for ${}^{9}$Be and ${}^{10}$Be.

\begin{figure}[h]
\begin{center}\includegraphics[width=200pt]{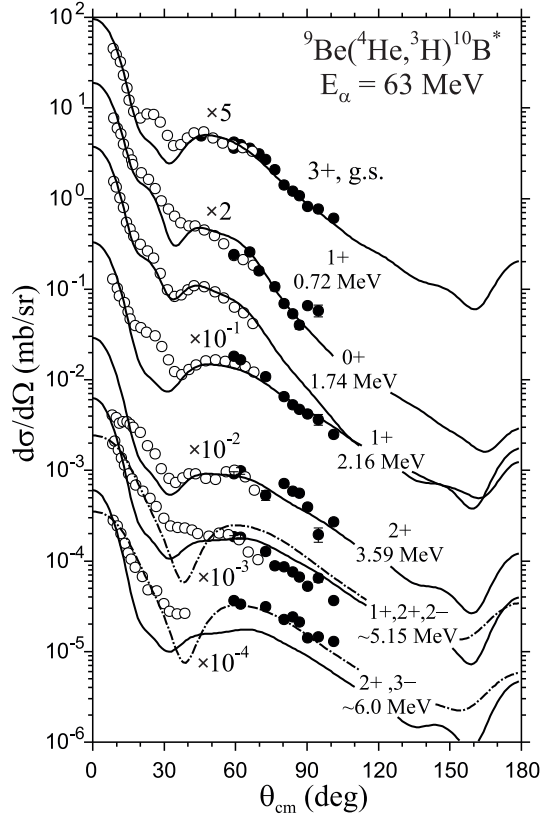}\end{center}
\caption{Angular distributions of the differential cross sections for the ground and low-lying excited states of ${}^{10}$B obtained in the reaction ${}^9$Be($\alpha$,t)${}^{10}$B: the 3$^+$ ground state of ${}^{10}$B, the first 1$^+$ level at 0.72 MeV, the sum of the cross sections for the 0$^+$ and 1$^+$ levels at 1.7 and 2.1 MeV, respectively, from the present experiment superimposed over the cross section of the 2.16 MeV level from Ref. [4], the 2$^+$ level at 3.6 MeV,the sum of the cross sections for the 2$^-$, 2$^+$ and 1$^+$ levels at about 5.1 MeV, and the 3$^-$ level at 6.1 MeV. Results of the present experiment are shown by solid symbols. Data from Ref.\cite{Ref4} are presented as open symbols.}\label{fig5}
\end{figure}

\subsection{${}^{10}$B} Differential cross sections versus cm-angles for the ground and low-lying excited states of ${}^{10}$B are plotted in Fig.\ref{fig5}. Results of the present experiment are shown by solid symbols, and data from Ref.\cite{Ref4} are presented as open symbols. DWBA calculations \cite{Ref23} for the ${}^{9}$Be($\alpha$,t)${}^{10}$B reaction were performed with the DWUCK5 code \cite{Ref24} with the fits to the differential cross sections for the ground and low-lying states given as thin solid lines in Fig.\ref{fig5}.

Fig.\ref{fig5} displays the observed angular distributions for the proton transfer reactions to the different ${}^{10}$B final states. One may notice a very good agreement between data obtained in Ref.\cite{Ref4} and our measurements. Theoretical results (solid and dash-dotted curves) fairly reproduce the data in case of well-defined final states. For the unresolved mixture of states at excitation energies of about 5.15 MeV and 6 MeV one may conclude that negative parity states corresponding to $l$ = 2 (shown by dash-dotted curves) provide better agreement with data than positive parity ones ($l$ = 1, solid curves).

Spectroscopic factors S$_f$ for the different states populated in the reaction ${}^{9}$Be(${}^{4}$He, t)${}^{10}$B are listed on the right side in Table \ref{Table3}. For the data corresponding to the mixture of a few levels an upper limit of spectroscopic factor was obtained, describing the data by one component only. S$_f$ values are in good agreement with those reported in the literature.

\subsection{Multiplet A = 10}
\begin{figure}[t]
\begin{center}\includegraphics[width=250pt]{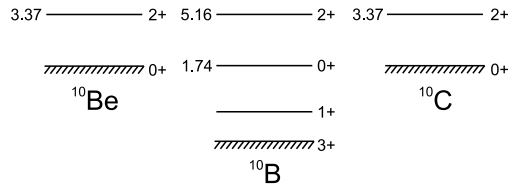}\end{center}
\caption{Level diagrams of low-lying states of ${}^{10}$Be, ${}^{10}$B and ${}^{10}$C, as members of the A = 10 multiplet. Values on the left sides correspond to the excitation energies in MeV, while numbers on the right side are spins and parities of corresponding states.}\label{fig6}
\end{figure}

The structure of ${}^{10}$Be, ${}^{10}$B and 10C nuclei was usually considered as two $\alpha$-clusters in the presence of two extra nucleons. Level diagrams for the low-energy excited states for these nuclei are shown in Fig.\ref{fig6}. One may see that the ${}^{10}$B ground state is shifted down by 2 MeV approximately. It may be treated as a three cluster configuration ${}^{10}$B = $\alpha$+d+$\alpha$ where the pairing of proton and neutron results in formation of a deuteron cluster inside. The 3$^+$ spin of this state also supports this assumption. In the case of 1.74 MeV excited state, it might be considered as a state, where the deuteron cluster becomes unbound. Thus it becomes four clusters configuration ${}^{10}$B(0$^+$, 1.74 MeV) = $\alpha$+p+n+$\alpha$ with uncorrelated proton and neutron. Two mirror ground states in ${}^{10}$Be and ${}^{10}$C in this case have to be of similar structure $\alpha$+N+N+$\alpha$. One of consequence of such an internal organization is the absence of the di-neutron component in the ${}^{10}$Be ground state wave function.

\begin{figure}[t]
\begin{center}\includegraphics[width=220pt]{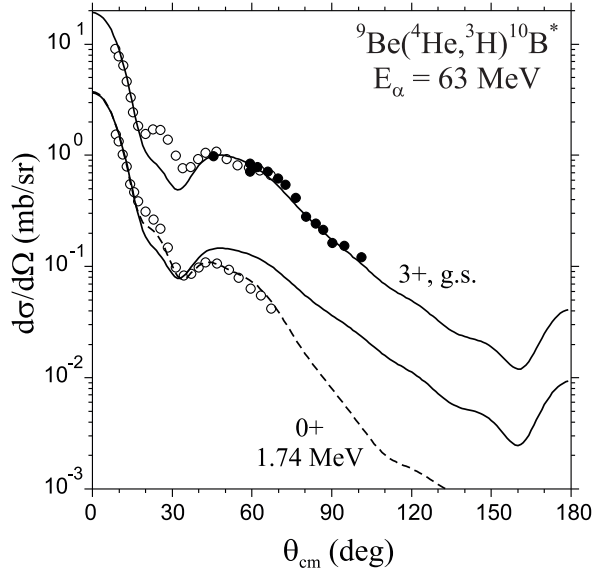}\end{center}
\caption{Differential cross sections for reaction ${}^{9}$Be(${}^{4}$He,t)${}^{10}$B leading to the ground and 1.74 MeV excited states of ${}^{10}$B nucleus. Solid symbols represent results of the present experiment, open ones is the data from Ref. \cite{Ref4}. Curves are the results of DWBA calculations (see text for explanations).}\label{fig7}
\end{figure}

Difference in the structure of the ground state and the 1.74 MeV state in ${}^{10}$B may also reveal itself in the difference of optical potentials for these exit channels. In Fig.\ref{fig7} the corresponding experimental data are compared with the results of DWBA calculations performed in the same manner as for the ${}^{3}$He + ${}^{10}$Be exit channel. Solid curves show theoretical cross sections obtained with the exit channel optical potential from Table \ref{Table3}. This potential was chosen on the basis of OM potential compilation form Ref.\cite{Ref28} with additional adjustment of parameters to the present data since Ref.\cite{Ref28} contains recommended optical potential for the lower energies. One may see quite good agreement between calculation and data on the case of ground-state channel. Applying the same OMP for the transfer to the 1.74 MeV state one gets the noticeable overestimation in the cross section at large angles. We found that in order to improve the agreement in the last case it is necessary to use the following parameters: $V_0$ = 85 MeV, $r_V$ = 1.14 fm and $W_0$ = 8 MeV. Corresponding result is shown in Fig.\ref{fig6} by the dashed curve and demonstrates excellent fit of the data. The obtained parameters turn out to be close to the OM potential for ${}^{3}$He + ${}^{10}$Be(g.s.) channel (see Table \ref{Table2}).

\section{Conclusions}
Angular distributions of the differential cross sections for the ${}^{9}$Be($\alpha$,$\alpha$)${}^{9}$Be$^*$, ${}^{9}$Be(${}^{4}$He,${}^{3}$He)${}^{10}$Be and ${}^{9}$Be(${}^{4}$He,t)${}^{10}$B reactions were measured. The observed states are most likely populated in one-step direct transfer reactions.

Experimental angular distributions for ground and a few low-lying states were described within the optical model and distorted-wave Born approximation frameworks. In the OM analysis and fits of the experimental data by CC calculations, it was found, generally speaking,  that the optical model parameters ($V_0$, $W_0$, radii and diffuseness) were not sensitive to the cluster structures of the excited states. To study cluster structure, a complicated experiment is planned in which decay of excited states by cluster emission will be investigated. However, to distinguish break-up into ${}^{4}$He and ${}^{5}$He will be not be a trivial kinematical problem.

The values 9/2$^-$ were assigned to the spin and parity of the 11.28 MeV state in ${}^{9}$Be. The obtained large $\beta_2$ value may be considered as confirmation of the cluster structure of the low-lying states of ${}^{9}$Be. However, it doesn't allow to give unambiguous preference to one of the possible configurations $\alpha$+$\alpha$+n or $\alpha$+${}^{5}$He. In order to improve the agreement between the theoretical prediction and the experimental data, related to this 9/2$^-$ state, an additional hexadecapole term $\beta_4$ in the definition of the ${}^{9}$Be radius had to be introduced.

With respect to ${}^{10}$Be,  it was found that in order to describe the data one needs a much smaller radius of the real part of the optical potential for the exit channel ($r_V$ = 0.95 fm) in comparison to the radius in the entrance channel ($r_V$ = 1.40 fm). This could be interpreted as evidence for the compactness of the ${}^{10}$Be nucleus.

The comparison of the angular distributions of the differential cross sections for the isobaric analog states of ${}^{10}$Be and ${}^{10}$B was done. The structure of ${}^{10}$Be, ${}^{10}$B and ${}^{10}$C nuclei was usually considered as two $\alpha$-clusters in the presence of two extra nucleons. One may see that the ${}^{10}$B ground state could be treated as a three cluster configuration ${}^{10}$B = $\alpha$+d+$\alpha$, where the pairing of proton and neutron results in formation of a deuteron cluster inside ${}^{10}$B. In the case of the 1.74 MeV excited state, it might be considered as a state where the deuteron cluster becomes unbound. Thus, it becomes a four-body configuration ${}^{10}$B(0$^+$, 1.74 MeV) = $\alpha$+p+n+$\alpha$, i.e. two $\alpha$-clusters with an uncorrelated proton and neutron pair.

Spectroscopic factors for the ground and excited states of ${}^{10}$B and ${}^{10}$Be were deduced. We found pretty good agreement between our results and the previous data.

\ack
We would like to thank the JYFL Accelerator Laboratory for giving us the opportunity to perform this study and the cyclotron staff for the excellent beam quality. This work was supported in part by the Russian Foundation for Basic Research (project numbers: 13-02-00533 and 13-07-00714), by grants to JINR (Dubna) of the Czech and Polish Republics, and by mobility grant from the Academy of Finland.

\section*{References}

\bibliographystyle{unsrt}
\bibliography{lukyanov}

\end{document}